\documentclass[Physsubmission]{SciPost}

\newcommand*{\cf}{cf.\ }
\newcommand*{\eg}{e.\,g.\ }
\newcommand*{\ie}{i.\,e.\ }

\hypersetup{
    colorlinks,
    linkcolor={red!50!black},
    citecolor={blue!50!black},
    urlcolor={blue!80!black}
}

\usepackage[bitstream-charter]{mathdesign}
\DeclareSymbolFont{usualmathcal}{OMS}{cmsy}{m}{n}
\DeclareSymbolFontAlphabet{\mathcal}{usualmathcal}

\begin{document}

\begin{center}{\Large \textbf{
Towards two- and three-loop QCD corrections to the width difference in $B_s-\bar{B}_s$ mixing\\
}}\end{center}

\begin{center}
Vladyslav Shtabovenko\textsuperscript{$\star$}
\end{center}

\begin{center}
Institut f{\"u}r Theoretische Teilchenphysik, \\
Karlsruhe Institute of Technology (KIT)
\\
76128 Karlsruhe, Germany \\

* v.shtabovenko@kit.edu
\end{center}

\begin{center}
\today
\end{center}

\definecolor{palegray}{gray}{0.95}
\begin{center}
\colorbox{palegray}{
  \begin{tabular}{rr}
  \begin{minipage}{0.1\textwidth}
    \includegraphics[width=35mm]{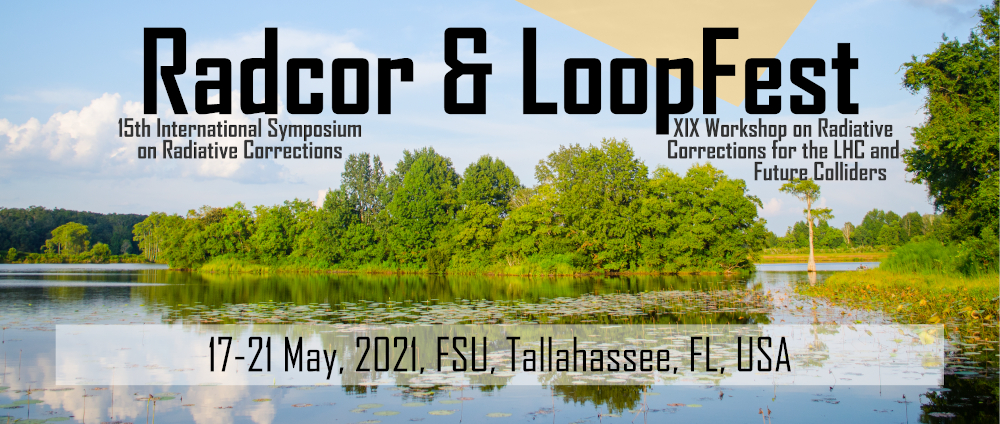}
  \end{minipage}
  &
  \begin{minipage}{0.85\textwidth}
    \begin{center}
    {\it 15th International Symposium on Radiative Corrections: \\Applications of Quantum Field Theory to Phenomenology,}\\
    {\it FSU, Tallahasse, FL, USA, 17-21 May 2021} \\
    \doi{10.21468/SciPostPhysProc.?}\\
    \end{center}
  \end{minipage}
\end{tabular}
}
\end{center}

\section*{Abstract}
{\bf
In this work we address the issue of large perturbative uncertainties 
in the theory prediction for $\Delta \Gamma_s$, the width difference in
the $B^0_s - \bar{B}^0_s$ mixing process. To this aim we complete 
important steps towards the full analytic result for the previously unknown Wilson coefficients from the matching  between $|\Delta B|=1$ and $|\Delta B|=2$ effective Hamiltonians at next-to-next-leading order (NNLO) in the strong coupling constant.
We provide a thorough discussion of technical and conceptual difficulties 
behind this computation and give an outlook regarding the availability
of the new NNLO theory prediction for $\Delta \Gamma_s$.
}

\vspace{10pt}
\noindent\rule{\textwidth}{1pt}
\tableofcontents\thispagestyle{fancy}
\noindent\rule{\textwidth}{1pt}
\vspace{10pt}

\section{Introduction}	
\label{sec:intro}

Oscillations of neutral meson systems such as $K^0 - \bar{K}^0$, $D^0 - \bar{D}^0$ and $B^0_q - \bar{B}^0_q$  with $q=s,d$
between their matter and antimatter states are an exciting manifestation of a genuine quantum mechanical phenomenon that can be experimentally observed in collider experiments. In the case of the $B^0_s - \bar{B}^0_s$ mixing there
are three relevant physical observables for which theoretical predictions can be confronted with
experimental measurements. These are $\Delta M_s$, $\Delta \Gamma_s$ and $a_{\textrm{fs}}^q$, which 
denote the oscillation frequency, the width difference and the flavor-specific CP-asymmetry respectively.
In this work our main interest is devoted to $\Delta \Gamma_s$, which is related to the imaginary part 
of the box diagrams describing the process $b \bar{s} \to \bar{b} s$. Owing to the absence of tree-level flavor changing neutral
currents (FCNCs) in the Standard Model (SM), we are dealing with a loop-induced process, where the
leading-order (LO) contribution starts at one loop. One of the motivations to consider  $\Delta \Gamma_s$ in more details concerns its role as a precision probe of the SM. In the Feynman diagrams relevant for this quantity possible new physics contributions may arise only in form of light beyond Standard Model
(BSM) particles that are weakly coupled to the SM sector. This is very different from $\Delta M_s$ which is on the contrary
extremely sensitive to hypothetical heavy particles contributing through loops. For this reason $\Delta \Gamma_s$ is universally regarded as an important indicator for our understanding of 
the SM flavor sector. As such, it has attracted wide interest from theory and experiment, where the latter \cite{LHCb:2019nin,CMS:2020efq,ATLAS:2020lbz} has already achieved the per-cent level accuracy \cite{HFLAV:2019otj} with  
\begin{equation}
	\Delta \Gamma^{\textrm{exp}}_s = (0.085 \pm 0.004) \textrm{ ps}^{-1}.
\end{equation}
Unfortunately, the present day situation on the theory side is less favorable.
The current most up-to-date theoretical predictions \cite{Beneke:1998sy,Ciuchini:2001vx,Ciuchini:2003ww,Lenz:2006hd,Asatrian:2017qaz,Asatrian:2020zxa} 
\begin{align}
	\Delta \Gamma_s^{\overline{\text{MS}}} &= (0.088 \pm 0.011{}_{\textrm{pert.}} \pm 0.002{}_{B,\tilde{B}_S} \pm 0.014_{\Lambda_{\textrm{QCD}}/m_b} ) \textrm{ ps}^{-1}, \\
	\Delta \Gamma_s^{\text{pole}} &= (0.077 \pm 0.015{}_{\textrm{pert.}} \pm 0.002{}_{B,\tilde{B}_S} \pm 0.017_{\Lambda_{\textrm{QCD}}/m_b} ) \textrm{ ps}^{-1}
\end{align}
exhibit large uncertainties from uncalculated QCD corrections to the Wilson coefficients (denoted as ``pert.'') that arise 
in the perturbative matching between $\mathcal{H}_{\textrm{eff}}^{|\Delta B|=1}$ and $\mathcal{H}_{\textrm{eff}}^{|\Delta B|=2}$
effective weak Hamiltonians at two and three loops. The determination of these corrections by means of an explicit analytic calculation is the main goal of this 
work. 

\section{Calculation}
\label{sec:calc}

On the $|\Delta B|=1$ side of the matching we employ the operator basis from \cite{Chetyrkin:1997gb} given by
\begin{align}
	\mathcal{H}_{\textrm{eff}}^{|\Delta B|=1} 
	&=   \frac{4G_F}{\sqrt{2}}  \left[
	-\, \lambda^s_t \Big( \sum_{i=1}^6 C_i Q_i + C_8 Q_8 \Big) 
	- \lambda^s_u \sum_{i=1}^2 C_i (Q_i - Q_i^u) \right. \\ \nonumber
	& \phantom{\frac{4G_F}{\sqrt{2}} \Big[}
	\left.
	+\, V_{us}^\ast V_{cb} \, \sum_{i=1}^2 C_i Q_i^{cu} 
	+ V_{cs}^\ast V_{ub} \, \sum_{i=1}^2 C_i Q_i^{uc} 
	\right]
	+ \mbox{h.c.},
\end{align}
with $\lambda^s_a = V_{as}^\ast V_{ab}$, where $V_{ij}$ denote CKM matrix elements and
$G_F$ stands for the Fermi constant. The Wilson coefficients $C_i$ arise from the matching
between SM diagrams and the $|\Delta B|=1$ effective theory, where all energy 
modes heavier than the $b$ quark mass $m_b$ are integrated out. We refer to \cite{Gerlach:2021xtb}
for the explicit definitions of the dimension-six operators $Q_i$. Of these, $Q_{1,2}$
are usually denoted as current operators, while $Q_{3-6,8}$ belong to the penguin category.
In addition to that, the operator basis also contains evanescent
operators $E[Q_i]$ \cite{Dugan:1990df,Herrlich:1994kh} whose matrix elements are of 
$\mathcal{O}(\varepsilon)$ (from $d = 4-2 \varepsilon$) reflecting the fact that certain 
Dirac algebra relations such as Fierz identities are ambiguous in dimensional regularization.

To obtain $\Delta \Gamma_s \approx 2 |\Gamma^s_{12}|$ we need to evaluate 
\begin{equation}
	\Gamma_{12}^s = \frac{1}{2 M_{B_s}} \,\mbox{Abs}\langle B_s|i\int d^4 x \,\, \mathcal{T}\,\,
	{\cal H}_{\rm eff}^{\Delta B=1}(x)
	{\cal H}_{\rm eff}^{\Delta B=1}(0)
	|\bar{B}_s\rangle\,,
\end{equation}
where $\mathcal{T}$ denotes time-ordering, $M_{B_s}$ stands for the meson mass and
``Abs'' specifies the absorptive part of the bilocal matrix element. The quantity $\Gamma_{12}^s$ can be decomposed into \cite{Beneke:1998sy}
\begin{eqnarray}
	\Gamma_{12}^s &=& - (\lambda_c^s)^2\Gamma^{cc}_{12} 
	- 2\lambda_c^s\lambda_u^s \Gamma_{12}^{uc} 
	- (\lambda_u^s)^2\Gamma^{uu}_{12} 
	\,,
\end{eqnarray}
Applying Heavy Quark Expansion (HQE) \cite{Khoze:1983yp,Shifman:1984wx,Khoze:1986fa,Chay:1990da,Bigi:1991ir,Bigi:1992su,Bigi:1993fe,Blok:1993va,Manohar:1993qn}
one can express $\Gamma_{12}^{ab}$ as
\begin{equation}
	\Gamma_{12}^{ab} 
	= \frac{G_F^2m_b^2}{24\pi M_{B_s}} \left[ 
	H^{ab}(z)   \langle B_s|Q|\bar{B}_s \rangle
	+ \widetilde{H}^{ab}_S(z)  \langle B_s|\widetilde{Q}_S|\bar{B}_s \rangle
	\right] 	+ \mathcal{O} (\Lambda_{\text{QCD}}/m_b), \label{eq:DB2}
\end{equation}
with $z \equiv m_c^2/m_b^2$ and $|\Delta B=2|$ operators
\begin{equation}
	Q = \bar{s}_i \gamma^\mu \,(1-\gamma^5)\, b_i \; \bar{s}_j \gamma_\mu
	\,(1-\gamma^5)\, b_j, \quad 
	\widetilde{Q}_S = \bar{s}_i \,(1-\gamma^5)\, b_j\; \bar{s}_j \,(1-\gamma^5)\,
	b_i, \label{eq:opDB2}
\end{equation}
where $i,j$ stand for the color indices of the fermions. In addition to that, in 
intermediate expressions one encounters color-switched versions of the 
operators given in Eq.\,\eqref{eq:opDB2} denoted as $\widetilde{Q}$ and $Q_S$. 
Notice that the NLO $|\Delta B|=2$ operator basis also includes evanescent operators. 
Last but not least, a particular linear combination 
of the operators $Q$, $Q_S$ and $\tilde{Q}_S$ yields a $1/m_b$-suppressed operator 
$R_0$ \cite{Beneke:1996gn,Beneke:1998sy} given by
\begin{equation}
	R_0 = \frac{1}{2} Q + Q_S + \widetilde{Q}_S + \mathcal{O}\left ( \frac{1}{m_b} \right ).
\end{equation}
Beyond LO the matrix element of $R_0$ develops corrections in $\alpha_s$ that would not
be $1/m_b$-suppressed unless $R_0$ is correctly renormalized. 
Further details regarding the $|\Delta B|=2$ effective operators are provided in \cite{Gerlach:2021xtb} and \cite{Gerlach:2021prep3L}.

In order to achieve the NNLO accuracy in the theory prediction for $\Delta\Gamma_s$ we
need to extend the knowledge of the Wilson coefficients $H^{ab}(z)$ and 
$\widetilde{H}^{ab}_S(z)$ to $\mathcal{O}(\alpha_s^2)$. This implies the evaluation
of all two-loop diagrams with two insertions of $|\Delta B|=1$ operators 
$Q_{1-6,8}$ and the computation of the current-current correlator
$Q_{1-2} \times Q_{1-2}$ at 3-loops. Notice that the two-loop contribution to 
$Q_{8} \times Q_{8}$ is actually $\mathcal{O}(\alpha_s^3)$, but in our framework
it does not pose any additional difficulties to obtain this result as a byproduct.
\begin{figure}[t]
	\begin{center}
		\begin{tabular}{cccc}
			\includegraphics[width=0.25\textwidth]{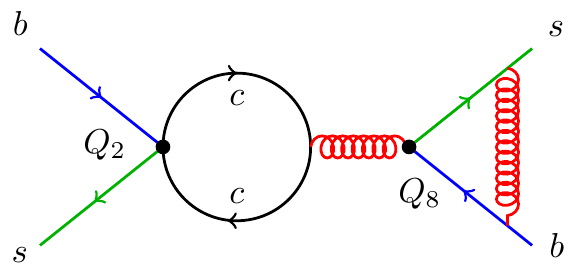} &
			\includegraphics[width=0.21\textwidth]{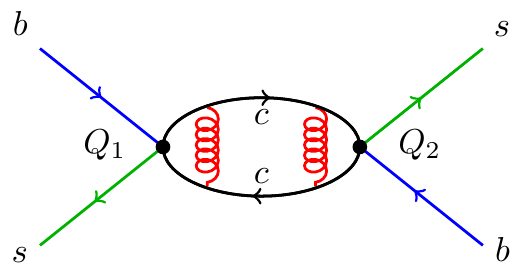} &
			\includegraphics[width=0.15\textwidth]{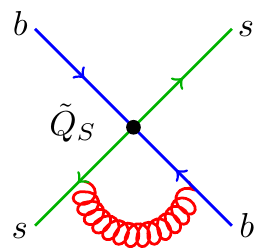} &
			\includegraphics[width=0.15\textwidth]{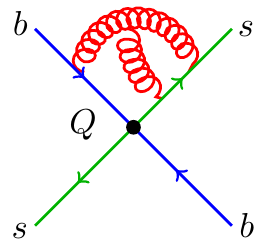} \\
			(a) & (b) & (c) & (d)
		\end{tabular}
	\end{center}
	\caption{\label{fig:sample}Sample $\Delta B=1$ and $\Delta B=2$ Feynman diagrams contributing to the
		process $b \bar{s} \to \bar{b} s$. Here (a) and (b) represent $\Delta B=1$ contributions
		from $Q_{2} \times Q_{8}$ (two loops) and $Q_{1} \times Q_{2}$ (three loops), while (c)
		and (d) show one- and two-loop matrix elements of $\Delta B=2$ operators $\tilde{Q}_s$ and $Q$.}
\end{figure}
All the above-mentioned diagrams on the $|\Delta B|=1$ side are then matched to 
one- and two-loop diagrams in the $|\Delta B|=2$ effective theory (\cf Fig.\,\ref{fig:sample}).
In the matching we put the $b$ quark external momenta on-shell, while setting the masses and external
momenta of the $s$ quarks to zero. The two-loop diagrams are expanded in the charm mass 
up to $\mathcal{O}(z)$, while at 3-loops we content ourselves with the $z=0$ limit.

In general, we employ dimensional regularization to handle both UV and IR divergences. However, 
as a cross check at two loops we also introduce a finite gluon mass $m_g$ as an IR regulator.
This is needed to ensure that all subtleties \cite{Ciuchini:2001vx} related to the treatment of evanescent operators
with $\varepsilon_{\text{UV}}=\varepsilon_{\text{IR}}$ (\eg keeping $\mathcal{O}(\varepsilon^2)$ and 
$\mathcal{O}(\varepsilon)$ contributions at LO and NLO for the NNLO matching) are properly taken into account. The full agreement between the two-loop matching coefficients obtained
from $m_g \neq 0$ and $m_g = 0$ calculations provides a nontrivial consistency check of our computations.

\section{Technical details}
\label{sec:tech}

Let us elaborate on the technical aspects behind this calculation.
The generation of diagrams with genuine 4-fermion operators always requires additional care
due to the relative signs arising from the corresponding vertices. These signs are usually not
handled by the diagram generator itself and need to be fixed by the code evaluating the corresponding
amplitudes. A popular way to sidestep this issue consists of introducing an auxiliary field (often called $\sigma$-particle
\cf \eg \cite{Nogueira:2006pq})
that connects two fermion-fermion-$\sigma$ vertices with each other. The Feynman rule for the 
propagator of this particle contains no denominator, while the numerator must be chosen such, that it reproduces
the original color structure of the corresponding 4-fermion operator. In this case the relative sings between
diagrams containing the so prepared 4-fermion operators automatically come out right.
In our calculation we decided to employ both methods which serves as a cross check for the
correctness of the generated amplitudes. 

For the purpose of creating diagrams with explicit 4-fermion vertices we implemented suitable
models in \textsc{FeynRules} \cite{Alloul:2013bka} and exported them to \textsc{FeynArts} \cite{Hahn:2000kx}. Notice 
that the inclusion of operators containing chains with more than 3 Dirac matrices requires some 
modifications of the \textsc{FeynRules} source code. By default, \textsc{FeynRules} chooses to
simplify such chains using the Chisholm identity for 3 Dirac matrices, which is highly undesirable in our calculation. This minor technical issue can be circumvented by commenting 
out the corresponding routine in the \texttt{Processing} section of the file \texttt{FeynArtsInterface.m}.
\textsc{FeynArts} does not fix the relative signs of 4-fermion operators
on its own, so in our case this task is handled by \textsc{FeynCalc} \cite{Mertig:1990an,Shtabovenko:2016sxi,Shtabovenko:2020gxv}. This allows us to generate representative diagrams at tree- and 1-loop level and compare the resulting amplitudes to the expressions
obtained using the $\sigma$-particle trick.

While \textsc{FeynArts} and \textsc{FeynCalc} are mainly employed for doing cross checks and reproducing
selected 1-loop results from the literature, the actual calculation relies on a \textsc{FORM}-based \cite{Kuipers:2012rf}
framework. In the first stage we generate the required diagrams with \textsc{QGRAF} \cite{Nogueira:1991ex} (using $\sigma$-particles)  and process them via the \textsc{C++} programs \textsc{q2e/exp} \cite{Harlander:1998cmq,Seidensticker:1999bb}. Here \textsc{q2e} is responsible for inserting Feynman rules into the output of \textsc{QGRAF} and extracting the form of the graph representing each of the generated diagrams. The latter information is needed for the graph-based topology identification
implemented in \textsc{exp}. 
As an alternative to \textsc{q2e} and the topology identification routines of \textsc{exp} we also employ
a new \textsc{Python}-based framework called \textsc{tapir} \cite{Gerlach:tapir}.
The resulting amplitudes are processed using the \textsc{FORM}-based in-house \textsc{calc} setup. The presence
of products of two Dirac chains resulting from the 4-fermion operators necessitates a dedicated treatment
of these algebraic structures. Here again we follow two different approaches that consist of constructing
a set of suitable Dirac and color projectors or performing tensor reductions. More details on the construction
of the projectors can be found in the appendix of \cite{Gerlach:2021xtb}.
For the generation of tensor reduction identities applicable to 2- and 3-loop integrals with one external momentum
and tensor ranks up to 10, we employ the \texttt{Tdec} function available in \textsc{FeynCalc}. Motivated by
the performance bottlenecks encountered during this project, the symmetrization procedure for tensor indices
implemented in \texttt{Tdec} has been rewritten from scratch using the algorithm from \cite{Pak:2011xt}. 
Furthermore, to speed up the process of solving the resulting symbolic linear equations, we augmented
the public collection of interfaces between \textsc{FeynCalc} and other HEP tools, known as \textsc{FeynHelpers}
\cite{Shtabovenko:2016whf}, with a link to \textsc{FERMAT} \cite{Lewis:FERMAT}, implemented as a new function called \texttt{FerSolve}. Using these publicly available tools we were able to generate and validate all the 
required tensor reduction rules and export them to \textsc{FORM}.
For the purpose of the IBP reduction \cite{Chetyrkin:1981qh,Tkachov:1981wb} we make use of \textsc{FIRE} 6 \cite{Smirnov:2019qkx}
and \textsc{LiteRed} \cite{Lee:2013mka}. Owing to the excellent performance of \textsc{FIRE} 6, the reduction itself does not pose any difficulties neither at two nor at three loops. 
For example, the reduction of about 59,000 single scale on-shell three-loop integrals appearing in our calculation can be performed in less than an hour on a laptop equipped
with an 8-core CPU (AMD Ryzen 4750U) and 32 GB of RAM.
\begin{figure}[t]
	\begin{center}
			\includegraphics[width=0.95\textwidth]{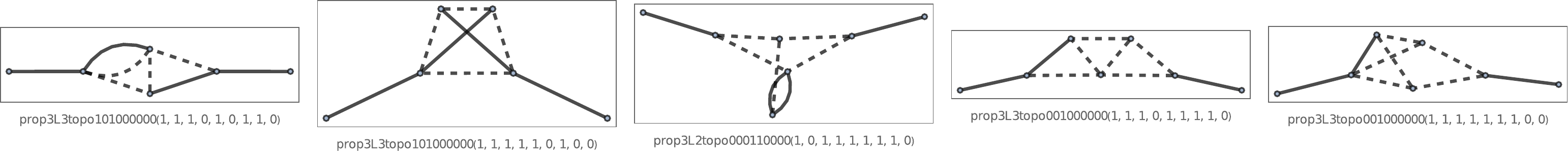}
	\end{center}
	\caption{Some of the nonfactorizing on-shell 3-loop master integrals occurring in our matching calculation.
		A solid line denotes a massive propagator with $m_b$, while a dashed line corresponds to a massless propagator.
		}
	\label{fig:masters}
\end{figure}
All master integrals encountered in this calculation are on-shell two-point functions or tadpoles.
The set of the master integrals at one- and two-loops resulting from diagrams with massless gluons
does not contain anything that is not already available in the literature (\cf \eg \cite{Smirnov:2012gma}). In particular,
analytic results for all propagator-type two-loop on-shell integrals with one mass scale can be found in \cite{Fleischer:1999hp}.

The three-loop on-shell integrals with one mass scale (\cf Fig.\,\ref{fig:masters}) turn out to be 
sufficiently simple to be integrated directly from the Feynman parametric form using 
\textsc{HyperInt} \cite{Panzer:2014caa}. To visualize the relevant integrals and derive  their Feynman parametrizations we made use of the new \textsc{FeynCalc} routines \texttt{FCLoopIntegralToGraph} and \texttt{FCFeynmanParametrize} that were specifically developed for this project. A more thorough description of these functions will be provided with the official release of \textsc{FeynCalc} 10 \cite{Shtabovenko:FCX}, although the routines themselves are already publicly available.
For most of our integrals the analytic results obtained with \textsc{HyperInt} involve complicated
Goncharov Polylogarithms (GPLs) \cite{Goncharov:1998kja} containing 6th root of unity that require further simplification. 
To this aim we employed
the packages \textsc{HyperLogProcedures} \cite{Schnetz:HLP} and \textsc{PolyLogTools} \cite{Duhr:2019tlz},
which allowed us to arrive at very simple and compact results, at least for the imaginary parts
relevant for our computation. The constants entering our expressions for the imaginary parts 
of the integrals are
\begin{equation}
	\pi, \, \ln(2), \, \zeta_2, \, \zeta_3, \, \zeta_4, \, \textrm{Cl}_2 \left  (\frac{\pi}{3} \right ), \, \sqrt{3}, \, \textrm{Li}_{4} \left ( \frac{1}{2} \right ), \, \ln \left ( \frac{1+\sqrt{5}}{2} \right ),
\end{equation}
with $\textrm{Cl}_2 (x) =  \frac{i}{2} \left ( \textrm{Li}_2 (e^{- ix}) - \textrm{Li}_2 (e^{ix}) \right )$. Explicit analytic results for all three-loop integrals entering the NNLO prediction for $\Delta \Gamma_s$
will be given in \cite{Gerlach:2021prep3L}. All the obtains results for master integrals have been validated 
numerically using \textsc{pySecDec} \cite{Borowka:2017idc,Borowka:2018goh,Heinrich:2021dbf} and
\textsc{FIESTA} \cite{Smirnov:2015mct}.

\section{Renormalization and matching}
\label{sec:ren}

The renormalization of the bare $|\Delta B|=1$ and  $|\Delta B|=2$ amplitudes
is done in the $\overline{\text{MS}}$ scheme. However, the QCD renormalization
constants alone are not sufficient to render the amplitudes UV finite. In addition
to that, one must also include the operator renormalization. In general, when renormalizing 
Wilson coefficients $W_i$ of our operators  we have
\begin{equation}
	(\vec{W}^{\textrm{bare}}, \vec{W}_E^{\textrm{bare}}) = (\vec{W}^{\textrm{ren}}, \vec{W}_E^{\textrm{ren}}) Z \equiv (\vec{W}^{\textrm{ren}}, \vec{W}_E^{\textrm{ren}})  \begin{pmatrix} Z_{QQ} & Z_{QE} \\ Z_{EQ} & Z_{EE} \end{pmatrix}.
\end{equation}
Here the submatrices $Z_{QQ}$ and $Z_{EE}$ describe the mixing of physical and evanescent operators among themselves, while 
the mixing of evanescent operators into the physical ones and vice versa is governed by $Z_{QE}$ and $Z_{EQ}$ respectively.
For the $|\Delta B|=1$ theory in the CMM basis, the NNLO renormalization matrix $Z$ has been computed in \cite{Gambino:2003zm}.
As far as the $|\Delta B|=2$ theory is concerned, we are not aware of a reference that provides full $Z$ including all four submatrices. For this reason we computed the required
$|\Delta B|=2$ renormalization matrix in a separate calculation, where the external momenta of $b$ and $s$ quarks were put off-shell and set to zero, while both quarks were given the same mass to regulate IR divergences. The so-obtained matrices for NLO and NNLO
will be given in \cite{Gerlach:2021prep2L} and \cite{Gerlach:2021prep3L}.

For calculations performed with a finite gluon mass, the UV-renormalization
renders the $|\Delta B|=1$ and $|\Delta B|=2$ amplitudes manifestly finite \ie free of $\varepsilon$-poles.
In the case of massless gluons the renormalized amplitudes still contain IR poles which, however,
cancel out in the matching as they should.

Upon putting everything together we obtain final contributions to the matching coefficients 
for all operator insertions at two loops at $\mathcal{O}(z)$ and preliminary (up to the finite renormalization of $R_0$) results for 
the three-loop correlator $Q_{1,2} \times Q_{1,2}$ at $\mathcal{O}(z^0)$. Regarding the two-loop 
results we observe full agreement between our $m_g=0$ and $m_g \neq 0$ calculations.
To put our work into perspective, let us briefly enumerate the existing results and 
explain how they are related to the corrections obtained here.

The two-loop $Q_{1,2} \times Q_{1,2}$ contribution with full $z$-dependence is known since many
years \cite{Beneke:1998sy} and was initially computed in the historical $|\Delta B|=1$ operator
basis and the old $|\Delta B|=2$ basis made of $Q$ and $Q_S$ operators. We reproduce this result
at $\mathcal{O}(z)$ and explicitly verify the correctness of the basis change formulas given
in \cite{Chetyrkin:1997gb} and \cite{Beneke:1998sy} for $|\Delta B|=1$ and $|\Delta B|=2$ theories respectively.
The same also applies to the one-loop $Q_{1,2} \times Q_{3-6}$ and $Q_{1,2} \times Q_{8}$
contributions from \cite{Beneke:1998sy}. We also reproduce the one-loop result for $Q_{3-6} \times Q_{3-6}$ from \cite{Beneke:1996gn}. 
The fermionic parts of the two-loop contributions $Q_{1,2} \times Q_{3-6}$ and $Q_{1,2} \times Q_{8}$
and of the one-loop contributions $Q_{3-6} \times Q_{8}$ and $Q_{8} \times Q_{8}$ with full $z$-dependence were computed in \cite{Asatrian:2020zxa}. The $N_f$-parts of our full matching coefficients agree with those results at $\mathcal{O}(z)$.
Finally, we also reproduce the $N_f$-piece of the three-loop $Q_{1,2} \times Q_{1-2}$ correlator given in
\cite{Asatrian:2017qaz} (at $\mathcal{O}(\sqrt{z})$) in the $z=0$ limit.

Therefore, apart from extending the existing incomplete results for 
$Q_{1,2} \times Q_{1-2}$ at three loops, $Q_{1,2} \times Q_{3-6}$ and $Q_{1,2} \times Q_{8}$ at two loops 
and $Q_{3-6} \times Q_{8}$ as well as $Q_{8} \times Q_{8}$  at one loop,
we also obtain new results for $Q_{3-6} \times Q_{8}$ and $Q_{8} \times Q_{8}$ at two loops. 
Since the explicit values of the matching coefficients are too lengthy to be presented here even for $N_c = 3$,
we refer to \cite{Gerlach:2021xtb} for the already published results for the two-loop contribution to
$Q_{1,2} \times Q_{3-6}$ and to \cite{Gerlach:2021prep2L,Gerlach:2021prep3L} for the remaining operator insertions.

In view of the still ongoing cross checks for the nonfermionic part of our three-loop result 
we are not yet in the position to provide an updated prediction for $\Delta\Gamma_s$.
However, we observe that as far as the two-loop contributions are concerned, the largest
relative shift of $\Delta\Gamma_s$ is generated by the $Q_{1,2} \times Q_{3-6}$ combination. Taking into account all current-current and current-penguin corrections up to 
$\mathcal{O}(\alpha_s)$ and penguin-penguin contributions up to 
$\mathcal{O}(\alpha_s^0)$, the ratio of the current-penguin contribution to
$\Delta\Gamma_s$ evaluates to
\begin{equation}
	\frac{\Delta\Gamma_s^{p,12\times36,\alpha_s^0}}{\Delta\Gamma_s}
	= 7.6 \% \quad ({\rm pole}),\quad 
	\frac{\Delta\Gamma_s^{p,12\times36,\alpha_s^0}}{\Delta\Gamma_s}
	= 6.1 \% \quad (\overline{\rm MS}), \label{eq:ratiQ_0}
\end{equation}
Including our new two-loop current-penguin correction both in the numerator
and denominator of Eq.\,\eqref{eq:ratiQ_0}  we find
\begin{equation}
	\frac{\Delta\Gamma_s^{p,12\times36,\alpha_s}}{\Delta\Gamma_s}
	= 0.3 \% \quad ({\rm pole}), \quad
	\frac{\Delta\Gamma_s^{p,12\times36,\alpha_s}}{\Delta\Gamma_s}
	= 1.4 \% \quad (\overline{\rm MS}).
\end{equation}
which implies a noticeable reduction of the existing perturbative uncertainties.
With ``$\overline{\text{MS}}$'' and ``pole'' we denote different ways to treat the $m_b^2$ prefactor
in Eq.\,\eqref{eq:DB2}. We can either evaluate it in the $\overline{\text{MS}}$ or in the on-shell
scheme. Notice, however, that even in the pole scheme all quantities except for the $m_b^2$ prefactor are handled in the $\overline{\text{MS}}$ scheme. More details on the numerical input parameters entering these comparisons can  be found
in \cite{Gerlach:2021xtb}. 

\section{Summary}
\label{sec:summary}

In order to reduce perturbative uncertainties entering theory predictions for the width 
difference $\Delta \Gamma_s$ in the  $B^0_s - \bar{B}^0_s$ mixing, we have calculated the previously
unknown QCD corrections to this quantity at two- and three-loop accuracy for all relevant combinations
of $|\Delta B|=1$ operator insertions. Our calculation is done in a fully analytic fashion including the evaluation
of new three-loop on-shell master integrals. For simplicity, we expand in the ratio $m_c^2/m_b^2$ up to
first order at two loops and set $m_c = 0$ at three loops.  In this limit we reproduce all already existing 
results from the literature. 
Since many of the literature results contain only fermionic
contributions, while our computation includes both $n_f$- and non-$n_f$ pieces, this work is
obviously instrumental to the task of improving theory predictions for  $\Delta \Gamma_s$.
A part of our new matching coefficients has already been published in
\cite{Gerlach:2021xtb} while the remaining two- and three-loop contributions will appear in 
\cite{Gerlach:2021prep2L} and \cite{Gerlach:2021prep3L} respectively.
For the future we also plan to increase the accuracy in $m_c^2/m_b^2$ by including higher orders
in this parameter or possibly by calculating the relevant master integrals with full $m_c$ dependence.

\section*{Acknowledgments}
We thank Artyom Hovhannisyan for useful discussions and for making his intermediate results from \cite{Asatrian:2017qaz}
available to us. We gratefully acknowledge the help of Erik Panzer, Oliver Schnetz, Gudrun Heinrich and Alexander Smirnov 
in making optimal use of the software tools \textsc{HyperInt}, \textsc{HyperLogProcedures}, \textsc{pySecDec} and \textsc{FIESTA} respectively. 

\paragraph{Funding information}
This research was supported by the Deutsche Forschungsgemeinschaft (DFG, German Research
 Foundation) under grant 396021762 — TRR 257 “Particle Physics Phenomenology after
the Higgs Discovery”. This report has been assigned preprint numbers TTP21-041 and P3H-21-079.

\bibliography{bmix-radcor.bib}

\nolinenumbers

\end{document}